\documentclass[letterpaper,11pt]{article}
\usepackage{amsmath,mathrsfs,amssymb,amsfonts,amsthm,mathtools}
\usepackage{physics,color,epsfig}
\usepackage{graphicx}
\usepackage{subfigure}
\usepackage{tabularx}
\usepackage{booktabs}
\usepackage[table]{xcolor}
\usepackage{tikz,tikz-cd}
\usetikzlibrary{shapes.geometric, arrows, positioning}
\usepackage{empheq}

\newcommand*\widefbox[1]{\fbox{\hspace{2em}#1\hspace{2em}}}
\pdfoutput=1
\usepackage{jheppub}
\usepackage[T1]{fontenc}

\newcommand{\nn}{\nonumber}

\hypersetup{
    colorlinks=true,
    linkcolor=blue,
    citecolor=blue,
    urlcolor=blue
}

\title{N-Photon Emission from Uniform Acceleration}

\author[a]{Arash Azizi}

\affiliation[a]{{\it The Institute for Quantum Science and Engineering,
Texas A\&M University,\\ College Station, TX 77843, U.S.A.}}

\emailAdd{sazizi@tamu.edu}

\abstract{
We present a generalized framework for $n$-photon processes involving a uniformly accelerated Unruh-DeWitt detector interacting with a massless scalar field. We utilize the $n^\text{th}$ order Dyson series to derive the final quantum state for an arbitrary number of interactions. Our analysis covers both even-order processes, which return the detector to its initial state, and odd-order processes, which result in a change of the detector's state. By employing a unified formalism and performing a complete, time-ordered integration, we obtain exact analytical expressions for the $n$-photon states. The results reveal a rich structure of resonant denominators corresponding to multi-particle processes, including new field-mediated resonances independent of the detector's energy gap for $n>2$. Crucially, the analysis of odd-order transitions reveals an exponential factor, $\exp(-\pi\omega/a)$, characteristic of the Unruh thermal bath. By considering processes starting from the detector's excited state, we demonstrate that the ratio of excitation to de-excitation amplitudes precisely recovers the Boltzmann factor, providing a higher-order confirmation of thermal detailed balance for the Unruh effect. This work provides a unified tool for studying multipartite entanglement and thermal phenomena in non-inertial frames.}

\begin{document}
\maketitle
\flushbottom

\section{Introduction}

The Unruh effect, a cornerstone of quantum field theory (QFT) in non-inertial frames, predicts that a uniformly accelerating observer perceives the Minkowski vacuum as a thermal bath \cite{Unruh1976, Fulling1973, Davies1975}. This phenomenon is often studied using the Unruh-DeWitt (UDW) detector model—a localized two-level quantum system coupled to a quantum field \cite{Unruh1976, Einstein100, Louko2008, Colosi2009Rovelli}. UDW detectors have proven invaluable for investigating relativistic quantum phenomena, such as entanglement generation in accelerated systems \cite{Reznik2003, Salton2015harvesting, Martin-Martinez2016harvesting, Bunney2023Circularmotion, Fewster2016Louko, Zhang2020harvesting, Barman2021harvesting, Liu2022harvesting} and quantum information tasks performed in curved spacetime contexts \cite{Su2014communication, Foo2020teleportation, Tjoa2022teleportation}.

While most studies focus on first-order processes, higher-order interactions provide a deeper window into the intricate structure of vacuum fluctuations and particle creation. Our recent work analyzed the two-photon emission from a single accelerating detector, demonstrating that the process can be understood as an Unruh-induced excitation followed by a Wigner-Weisskopf-like radiative decay \cite{Svidzinsky2021PRL, Svidzinsky2021PRR, Scully2022, Azizi2025TPE}. This second-order process revealed how the detector can harvest bipartite entanglement from the vacuum. The present paper extends these investigations from the second order to the general $n^\text{th}$ order, providing a comprehensive framework for multi-particle phenomena in accelerated frames.

Our primary contribution is a complete and explicit calculation of the $n^\text{th}$ order Dyson series for this process. We derive exact analytical formulas for the final $n$-photon state for both even-order ($g \to g$) and odd-order ($g \to e$) transitions. By fully evaluating the time-ordered integrals, we reveal a rich resonant structure. Our key findings are:
\begin{enumerate}
    \item The discovery of new ``field-mediated'' resonances for $n>2$ that are independent of the detector's energy gap,
    \item An explicit demonstration that odd-order excitation processes are suppressed by the characteristic Unruh thermal factor $e^{-\pi\omega/a}$,
     \item A confirmation of thermal detailed balance by showing that the ratio of de-excitation to excitation probabilities matches $e^{2\pi\omega/a}$, confirming the Boltzmann factor for detailed balance.
\end{enumerate}
Our results demonstrate how an accelerating detector can catalyze the creation of multipartite entangled states from the vacuum, with correlations dictated by the geometry of the accelerated frame. The generation of such non-local entanglement from a localized interaction highlights the underlying structure of the Minkowski vacuum itself \cite{Higuchi1992, Lin2006, Raval1996}. These findings not only deepen our understanding of observer-dependent phenomena in QFT but may also inform the development of relativistic quantum information protocols, such as vacuum-induced quantum gates \cite{Azizi2025VIQG}, in non-inertial frames.

\section{Theoretical Framework} \label{sec:framework}

\subsection{Field Quantization and Unified Formalism}

We begin by establishing the framework for a massless scalar field \(\Phi\) in \(1{+}1\)-dimensional Minkowski spacetime, interacting with a uniformly accelerated Unruh–DeWitt detector. The detector is a two-level system with states \(\ket{g}\) and \(\ket{e}\) separated by an energy gap \(\omega\). In the detector’s proper time \(\tau\), the interaction Hamiltonian is
\begin{equation}
H_{\text{int}}(\tau)
= g\,\partial_{\tau}\Phi\!\big(x(\tau)\big)\,\Big(\sigma^{\dagger}e^{i\omega\tau}+\sigma e^{-i\omega\tau}\Big),
\label{H_int}
\end{equation}
where \(g\) is the coupling constant and \(\sigma,\sigma^{\dagger}\) are the detector ladder operators. The proper-time derivative coupling is chosen for both physical and mathematical reasons. Physically, it mirrors a dipole-like response whereby transitions are governed by the local time rate of change of the field evaluated on the detector’s worldline. Mathematically, in \(1{+}1\) dimensions for a massless field it regularizes the pulled-back Wightman distribution and yields well-defined transition rates without auxiliary infrared regulators. Moreover, the chain rule along the trajectory implies
\[
du\,\partial_{u}=d\tau\,\partial_{\tau},\qquad dv\,\partial_{v}=d\tau\,\partial_{\tau},
\]
so the Jacobian from light-cone to proper-time coordinates cancels, and the Dyson series may be written directly as proper-time integrals. (Replacing \(\partial_{\tau}\Phi\) by \(\Phi\) removes only vertex-frequency prefactors while leaving the time-ordering structure and selection rules unchanged.)

To analyze emissions into right-traveling waves (RTW, coordinate $u=t-x$) and left-traveling waves (LTW, coordinate $v=t+x$), it is advantageous to employ a unified formalism using a direction parameter $\chi = \pm 1$. We define a single light-cone coordinate $w = t - \chi x$, where $\chi=1$ corresponds to $u$ and $\chi=-1$ corresponds to $v$. The field operator $\Phi_\chi(w)$ is expanded in Unruh modes \cite{Unruh1976, UnruhWald1984} as:
\begin{align}
\Phi_{\chi}(w)= \int_{-\infty}^{+\infty} d\Omega
\Bigg\{& \Big(
\theta(w) f(\Omega) w^{i\Omega}
+ \theta(-w) f(-\Omega) (-w)^{i\Omega} \Big) A_{\chi,\Omega} \nn \\
&+ \Big( \theta(w) f(\Omega) w^{-i\Omega}
+ \theta(-w) f(-\Omega) (-w)^{-i\Omega} \Big) A_{\chi,\Omega}^{\dagger}
\Bigg\},  \label{Unruh.mode}
\end{align}
where $A_{1,\Omega} \equiv A_\Omega$ (RTW annihilation operator) and $A_{-1,\Omega} \equiv B_\Omega$ (LTW annihilation operator). They satisfy $[A_{\chi,\Omega}, A_{\chi',\Omega'}^\dagger] = \delta_{\chi\chi'} \delta(\Omega - \Omega')$. The normalization factor is given by
\begin{equation}
f(\Omega) = \frac{e^{- \pi \Omega / 2}}{\sqrt{8\pi |\Omega| \sinh(\pi |\Omega|)}}.\label{f}
\end{equation}
The detector follows a trajectory of constant proper acceleration $a$ in the right Rindler wedge ($u < 0, v > 0$), given by $u(\tau) = -(1/a)e^{-a\tau}$ and $v(\tau) = (1/a)e^{a\tau}$. In the unified coordinate, this is $-\chi a w = e^{-\chi a \tau}$. Substituting this into the field expansion, the field along the detector's worldline becomes:
\begin{align}
\Phi_{\chi}(\tau)
&= \int_{-\infty}^{+\infty} d\Omega f(-\chi\Omega)\Big(
a^{-i\Omega} e^{-i \chi a \Omega \tau} A_{\chi,\Omega}
+ a^{i\Omega} e^{i \chi a \Omega \tau} A_{\chi,\Omega}^\dagger
\Big). \label{Unruh.mode.chi}
\end{align}
Since we are interested in photon emission from the Minkowski vacuum $\ket{0_M}$, only the creation operator terms $A_{\chi,\Omega}^\dagger$ will contribute.

\subsection{The Two-Photon Process Revisited}

Before tackling the general $n$-photon case, we briefly review the second-order, two-photon emission process using the unified formalism. The detector starts in $\ket{g}$ and returns to $\ket{g}$, so the process is described by the second-order term in the Dyson series. The final state is a superposition of four channels, $\ket{\Psi_f} = \sum_{\chi_1, \chi_2} \ket{\Psi_f}_{\chi_1 \chi_2}$, corresponding to RR, LL, RL, and LR emissions. A single channel is given by:
\begin{align}
\ket{\Psi_f^{(2)}}_{\chi_1\chi_2}
= \frac{i g^2}{4\hbar^2} \int_{-\infty}^{+\infty} d\Omega
\frac{\Omega}{\left(\frac{\omega}{a} - \chi_1 \Omega\right)}
\frac{a^{i\Omega(1-\chi_1\chi_2)}}{\sinh(\pi\Omega)}
A^\dagger_{\chi_1,\Omega}
A^\dagger_{\chi_2, -\chi_1\chi_2\Omega}
\ket{0_M}\ket{g}\,. \label{psi_f_2_final}
\end{align}
This result, derived in detail in \cite{Azizi2025TPE}, shows that the two emitted photons have correlated Unruh frequencies. The resonant denominator $(\omega/a - \chi_1\Omega)$ clearly shows the process is peaked when the Unruh frequency matches the detector gap, a hallmark of the Unruh effect.

\section{\texorpdfstring{$n^\text{th}$ Order Emission Processes}{}}

\subsection{Even-Order Processes: \texorpdfstring{ $g \to g$}{}}

We first generalize to an $n$-photon emission where $n$ is even. This process returns the detector to its ground state. The $n^\text{th}$ order contribution to the final state is given by the time-ordered integral:
\begin{equation}
\ket{\Psi_f^{(n)}} = \left(\frac{-i}{\hbar}\right)^n \int_{-\infty}^{\infty} d\tau_1 \dots \int_{-\infty}^{\tau_{n-1}} d\tau_n H_{\text{int}}(\tau_1) \dots H_{\text{int}}(\tau_n) \ket{0_M} \ket{g}. \label{eq:dyson_n_even}
\end{equation}
Consistent with $\tau_n \le \cdots \le \tau_2 \le \tau_1$, the detector operators alternate so that the earliest vertex at $\tau_n$ carries $\sigma^\dagger e^{+i\omega\tau_n}$, and the latest at $\tau_1$ carries $\sigma e^{-i\omega\tau_1}$, ensuring $\ket{g}\!\to\!\ket{e}\!\to\!\cdots\!\to\!\ket{g}$. For a specific channel $\{\chi_j\}$, the state is:
\begin{align}
\ket{\Psi_f^{(n)}}_{\{\chi_j\}} = \left(\frac{-ig}{\hbar}\right)^n
    &\int_{-\infty}^{+\infty} d\tau_1 \dots \int_{-\infty}^{\tau_{n-1}} d\tau_n \left( \frac{\partial \Phi_{\chi_1}^{(-)}(\tau_1)}{\partial \tau_1} e^{-i\omega\tau_1} \right)
    \left( \frac{\partial \Phi_{\chi_2}^{(-)}(\tau_2)}{\partial \tau_2} e^{i\omega\tau_2} \right) \dots \nonumber \\
    &\quad \times \left( \frac{\partial \Phi_{\chi_n}^{(-)}(\tau_n)}{\partial \tau_n} e^{i\omega\tau_n} \right)
    \ket{g} \ket{0_M},
\end{align}
where $\Phi^{(-)}$ denotes the negative frequency contribution of the field which contains the creation operator. Substituting the field expansion (Eq. \ref{Unruh.mode.chi}) and performing the derivatives $\partial/\partial\tau_j$ brings down a factor of $ia\chi_j\Omega_j$. The expression becomes a series of nested integrals over both time and frequency.
\begin{align}
\ket{\Psi_f^{(n)}} = \left(\frac{ga}{\hbar}\right)^n \int d\Omega_1 \dots d\Omega_n
&\left(\prod_{j=1}^n \chi_j\Omega_j f(-\chi_j\Omega_j) a^{i\Omega_j} A^\dagger_{\chi_j,\Omega_j}\right) \nonumber \\
\times &\int_{-\infty}^\infty d\tau_1 e^{iE_1\tau_1} \int_{-\infty}^{\tau_1} d\tau_2 e^{iE_2\tau_2} \dots \int_{-\infty}^{\tau_{n-1}} d\tau_n e^{iE_n\tau_n},
\end{align}
where the energy term for each vertex is $E_j = a\chi_j\Omega_j + (-1)^j \omega$, for $j=1,\dots,n$.

We understand all time integrals with the standard adiabatic regulator $e^{-\epsilon|\tau|}$, $\epsilon\to 0^+$, which produces the $-i \epsilon$ prescription in all energy denominators. Repeating this procedure recursively for all $n-1$ nested integrals, we obtain a product of energy denominators:
\begin{equation}
\int_{-\infty}^{\infty} d\tau_1 \dots \int_{-\infty}^{\tau_{n-1}} d\tau_n \left(\prod_{j=1}^n e^{iE_j\tau_j}\right)
= \frac{2\pi\,\delta\!\big(\sum_{j=1}^n E_j\big)}{i^{\,n-1}\prod_{k=2}^{n} \big(\sum_{j=k}^{n} E_j -i \epsilon\big)}. 
\end{equation}
Since $n$ is even, the detector energy terms cancel in the delta function, $\sum (-1)^j \omega = 0$, and the conservation law simplifies to $a\sum_{j=1}^n \chi_j\Omega_j = 0$.

The overall amplitude of the $n$-photon process includes a product of the Unruh mode normalization factors. We can evaluate this term by substituting the definition of $f(\Omega)$ from Eq. \eqref{f}:
\[
\prod_{i=1}^n f(-\chi_i \Omega_i) = \prod_{i=1}^n \frac{e^{\frac{\pi}{2} \chi_i \Omega_i}}{\sqrt{8\pi(-\chi_i\Omega_i)\sinh(-\pi\chi_i\Omega_i)}} = \frac{e^{\frac{\pi}{2} \sum_{i=1}^n \chi_i \Omega_i}}{(8\pi)^{n/2} \prod_{i=1}^n \sqrt{-\chi_i\Omega_i\sinh(-\pi\chi_i\Omega_i)}}.
\]
The expression in the denominator can be simplified by using the identity $-\Omega\sinh(-\pi\Omega) = \Omega\sinh(\pi\Omega)$. Applying this to each term in the product, we obtain the simplified result:
\begin{align}
    \prod_{i=1}^n f(-\chi_i \Omega_i)
= \frac{e^{\frac{\pi}{2} \sum_{i=1}^n \chi_i \Omega_i}}{(8\pi)^{n/2} \sqrt{\prod_{i=1}^n |\Omega_i| \,\sinh(\pi|\Omega_i|)}}\,.
\end{align}
This expression reveals how the thermal factor, contained in the exponential, is separated from the rest of the mode normalization. For even-$n$ processes where $\sum \chi_i\Omega_i = 0$, this exponential term becomes unity.

Performing the $n-1$ nested time integrals with the adiabatic regulator $e^{-\epsilon|\tau|}$ ($\epsilon\to 0^+$) gives
\begin{equation}
\int_{-\infty}^{\infty} d\tau_1 \dots \int_{-\infty}^{\tau_{n-1}} d\tau_n \Big(\prod_{j=1}^n e^{iE_j\tau_j}\Big)
= \frac{2\pi\,\delta\!\big(\sum_{j=1}^n E_j\big)}{i^{\,n-1}\prod_{k=2}^{n} \big(\sum_{j=k}^{n} E_j -i \epsilon\big)}.
\end{equation}
Since $n$ is even, $\sum_{j=1}^n (-1)^j \omega = 0$, so the conservation law reduces to $a\sum_{j=1}^n \chi_j\Omega_j=0$, i.e., $\sum_{j=1}^n \chi_j\Omega_j=0$ after using $\delta(aX)=\delta(X)/a$. Using the identity $-\Omega \sinh(-\pi\Omega)=\Omega \sinh(\pi\Omega)$, the Unruh normalization product becomes
\[
\prod_{i=1}^n f(-\chi_i \Omega_i)
= \frac{e^{\frac{\pi}{2} \sum_{i=1}^n \chi_i \Omega_i}}{(8\pi)^{n/2} \sqrt{\prod_{i=1}^n |\Omega_i|\,\sinh(\pi|\Omega_i|)}}.
\]
For even $n$ the exponential factor equals unity on the support of the delta function. Eliminating $\Omega_1$ with the delta function we obtain
\begin{empheq}[box=\widefbox]{align}
\label{eq:n_photon_final_even_fixed}
\ket{\Psi_f^{(n)}}_{\{\chi_j\}}
= C_n \!\int d\Omega_2 \cdots d\Omega_n\;
&\Bigg( \prod_{j=1}^n \frac{\chi_j \Omega_j\, a^{i\Omega_j}}{\sqrt{|\Omega_j|\sinh(\pi|\Omega_j|)}}\,
A^\dagger_{\chi_j,\Omega_j} \Bigg)_{\Omega_1 = -\chi_1\sum_{k=2}^n \chi_k\Omega_k} \\
&\times \Bigg[\prod_{k=2}^n \frac{1}{a \sum_{j=k}^n \chi_j \Omega_j \;+\; \omega \sum_{j=k}^n (-1)^j -i \epsilon}\Bigg]\ket{0_M}\ket{g}, \nonumber
\end{empheq}
where
\begin{align}
  C_n = \frac{2\pi\,i^{\,1-n}}{(8\pi)^{n/2}}\left(\frac{g}{\hbar}\right)^{n} a^{n-1}.  
\end{align}

\subsection{Odd-Order Processes: \texorpdfstring{$g \to e$}{}}

We now take $n$ odd, so the detector ends in the excited state $\ket{e}$. With $\tau_n\le\cdots\le \tau_2\le \tau_1$, the earliest vertex at $\tau_n$ carries $\sigma^\dagger e^{+i\omega\tau_n}$ and the latest at $\tau_1$ carries $\sigma^\dagger e^{+i\omega\tau_1}$, with alternating $\sigma$ insertions in between, yielding the net transition $\ket{g}\!\to\!\ket{e}$.

 As before, the Unruh-mode annihilation operators satisfy $A_{\Omega}\ket{0_M}=0$, $B_{\Omega}\ket{0_M}=0$, and $[A_{\chi,\Omega},A_{\chi',\Omega'}^\dagger]=\delta_{\chi\chi'}\delta(\Omega-\Omega')$.

The $n^\text{th}$-order state reads
\begin{align}
\ket{\Psi_f^{(n)}} = \left(\frac{-ig}{\hbar}\right)^{\!n}
&\int_{-\infty}^{+\infty} d\tau_1\, \frac{\partial\Phi^{(-)}_{\chi_1}}{\partial\tau_1}\, e^{+i\omega\tau_1}\,\sigma^\dagger
\int_{-\infty}^{\tau_1} d\tau_2\, \frac{\partial\Phi^{(-)}_{\chi_2}}{\partial\tau_2}\, e^{-i\omega\tau_2}\,\sigma \,\cdots \nonumber\\
&\cdots\int_{-\infty}^{\tau_{n-1}} d\tau_n\, \frac{\partial\Phi^{(-)}_{\chi_n}}{\partial\tau_n}\, e^{+i\omega\tau_n}\,\sigma^\dagger \,\ket{0_M}\ket{g},
\end{align}
where each proper-time derivative contributes a factor $i\,a\,\chi_j\Omega_j$ upon inserting the field expansion; the net $a^n$ from the vertices will combine with the $1/a$ from the delta function below to give the overall $a^{n-1}$ scaling. The recursive time integrals obey
\[
\int_{-\infty}^{\tau_{n-1}} d\tau_n\, e^{iE_n\tau_n}
=\frac{e^{iE_n\tau_{n-1}}}{i(E_n-i \epsilon)}\,,
\qquad
\int_{-\infty}^{\tau_{n-2}} d\tau_{n-1}\, e^{i(E_{n-1}+E_n)\tau_{n-1}}
=\frac{e^{i(E_{n-1}+E_n)\tau_{n-2}}}{i(E_{n-1}+E_n-i \epsilon)}\,,
\]
and repeating to the top yields
\begin{equation}
\int_{-\infty}^{\infty}\! d\tau_1 \dots \int_{-\infty}^{\tau_{n-1}}\! d\tau_n \Big(\prod_{j=1}^{n} e^{iE_j\tau_j}\Big)
= \frac{2\pi\,\delta\!\Big(\sum_{j=1}^{n}E_j\Big)}{i^{\,n-1}\prod_{k=2}^{n}\big(\sum_{j=k}^{n}E_j-i \epsilon\big)}.
\end{equation}
For odd $n$, the vertex energies are $E_j=a\,\chi_j\Omega_j+(-1)^{j+1}\omega$, so $\sum_{j=1}^{n}(-1)^{j+1}\omega=+\omega$ and the conservation law becomes
\begin{equation}
a\sum_{j=1}^{n}\chi_j\Omega_j+\omega=0,
\end{equation}
which fixes $\sum_{j=1}^{n}\chi_j\Omega_j=-\omega/a$.

The Unruh-mode normalization product isolates a thermal factor:
\[
\prod_{i=1}^n f(-\chi_i \Omega_i)
= \prod_{i=1}^n \frac{e^{\frac{\pi}{2}\chi_i\Omega_i}}{\sqrt{8\pi\,|\Omega_i|\,\sinh(\pi|\Omega_i|)}}
= \frac{e^{\frac{\pi}{2}\sum_{i=1}^{n}\chi_i\Omega_i}}{(8\pi)^{n/2}\sqrt{\prod_{i=1}^{n}|\Omega_i|\,\sinh(\pi|\Omega_i|)}}\,.
\]
Using $\sum_{i}\chi_i\Omega_i=-\omega/a$ gives the amplitude-level factor $e^{-\pi\omega/(2a)}$, and hence a probability-level Unruh suppression $e^{-\pi\omega/a}$ after squaring.

Eliminating $\Omega_1$ with the delta function, the odd-$n$ final state in a specific channel $\{\chi_j\}$ takes the form
\begin{empheq}[box=\widefbox]{align}
\label{eq:n_photon_final_odd}
\ket{\Psi_f^{(n)}}_{\{\chi_j\}}
= C'_n\, e^{-\frac{\pi\omega}{2a}}
\!\int d\Omega_2 \cdots d\Omega_n\;
&\Bigg( \prod_{j=1}^n \frac{\chi_j\Omega_j\, a^{i\Omega_j}}{\sqrt{|\Omega_j|\sinh(\pi|\Omega_j|)}}\,
A^\dagger_{\chi_j, \Omega_j} \Bigg)_{\Omega_1=-\chi_1\left(\frac{\omega}{a} + \sum_{i=2}^n \chi_i\Omega_i\right)} \\
&\times \Bigg[ \prod_{k=2}^n \frac{1}{a \sum_{j=k}^n \chi_j \Omega_j + \omega \sum_{j=k}^n (-1)^{j+1} -i \epsilon} \Bigg]
\ket{0_M}\ket{e}, \nonumber
\end{empheq}
with
\[
C'_n=\frac{2\pi\,i^{\,1-n}}{(8\pi)^{n/2}}\left(\frac{g}{\hbar}\right)^{\!n} a^{\,n-1}(-1)^{\frac{n-1}{2}},
\qquad
\Omega_1=-\chi_1\!\left(\frac{\omega}{a}+\sum_{i=2}^{n}\chi_i\Omega_i\right).
\]
The key physical result is that any process that excites the detector is exponentially suppressed by a factor set by the ratio of the gap to the acceleration, with the amplitude acquiring $e^{-\pi\omega/(2a)}$ and the transition probability $e^{-\pi\omega/a}$.

\subsection{Initial State Dependence and Thermal Detailed Balance}

The thermal nature of the Unruh effect can be further confirmed by considering processes that begin with the detector in its excited state, $\ket{e}$. Let's analyze how the calculation changes. If the initial state is $\ket{e}$, the first interaction at the earliest time $\tau_n$ must be a de-excitation, involving the operator $\sigma e^{-i\omega\tau_n}$. The subsequent operators must then alternate to return the detector to a final state.

A careful analysis of the operator ordering shows that the amplitude for any $n^\text{th}$ order process starting from $\ket{e}$ is identical to the corresponding process starting from $\ket{g}$, but with the simple substitution $\omega \to -\omega$.

This substitution does not dramatically alter the picture for even $n$ processes (which map $\ket{e} \to \ket{e}$), as $\omega$ only appears in the denominators. However, it drastically changes the physics for odd-$n$ transitions, which connect the ground and excited states.
\begin{enumerate}
    \item \textbf{Excitation ($g \to e$, $n$ odd):} As shown in Eq. \eqref{eq:n_photon_final_odd}, the amplitude for this process contains the exponential factor $e^{-\pi\omega/2a}$. This represents an Unruh-like thermal excitation, which is suppressed for large energy gaps or small accelerations.

    \item \textbf{De-excitation ($e \to g$, $n$ odd):} For this process, we start with $\ket{e}$ and make the substitution $\omega \to -\omega$. The exponential factor in the amplitude becomes $e^{-\pi(-\omega)/2a} = e^{+\pi\omega/2a}$. This corresponds to stimulated and spontaneous emission, which is exponentially enhanced.
\end{enumerate}
This relationship is illustrated in Fig. \ref{fig:detailed_balance}.
\begin{figure}[ht]
\centering
\begin{tikzcd}
    \ket{g} \arrow[r, "e^{-\frac{\pi\omega}{a}}"] & \ket{e} & \quad & \ket{e} \arrow[r, "e^{+\frac{\pi\omega}{a}}"] & \ket{g}
\end{tikzcd}
\caption{A diagram illustrating the exponential factors governing the transition rates for excitation ($\ket{g} \to \ket{e}$) and de-excitation ($\ket{e} \to \ket{g}$) processes. The excitation is exponentially suppressed, while de-excitation is enhanced, consistent with thermal detailed balance.}
\label{fig:detailed_balance}
\end{figure}
The ratio of the probabilities for these two processes is therefore:
\begin{equation}
    \frac{P(e \to g)}{P(g \to e)} = \frac{|\mathcal{A}(e \to g)|^2}{|\mathcal{A}(g \to e)|^2} \propto \frac{|e^{+\pi\omega/2a}|^2}{|e^{-\pi\omega/2a}|^2} = \frac{e^{+\pi\omega/a}}{e^{-\pi\omega/a}} = e^{2\pi\omega/a}.
\end{equation}
This is precisely the principle of detailed balance for a two-level system in a thermal bath at temperature $T_U = a/2\pi k_B$. The ability of our $n^\text{th}$ order formalism to recover this fundamental result provides a powerful, non-perturbative confirmation of the thermal character of the quantum vacuum as seen by an accelerating observer.

\section{\texorpdfstring{Analysis of $n$-photon Processes}{}}

The $n$-photon states derived in the previous sections contain a rich physical structure, particularly within the energy denominators of the amplitudes. The processes are sharply peaked when these denominators approach zero, revealing the resonant conditions for multi-photon emission. Furthermore, the overall energy conservation, enforced by the Dirac delta function, dictates both the correlations between the emitted photons and the thermal character of the interaction.

\subsection{Resonance Structure}

The resonant behavior is governed by the product of denominators, $\mathcal{D}$, which takes a different form for even and odd-$n$ processes:
\begin{align}
    \mathcal{D}_{\text{even}} =& \prod_{k=2}^n \left( a \sum_{j=k}^n \chi_j \Omega_j + \omega \sum_{j=k}^n (-1)^j  \right), \nn\\
    \mathcal{D}_{\text{odd}} =& \prod_{k=2}^n \left( a \sum_{j=k}^n \chi_j \Omega_j + \omega \sum_{j=k}^n (-1)^{j+1}  \right).
\end{align}
Analyzing the structure of these products reveals two distinct classes of resonances.

\subsubsection*{Detector-Mediated Resonances}
This class of resonance involves denominators containing the detector's energy gap, $\omega$. They correspond to points in the multi-step process where the detector's internal state changes, directly mediating energy transfer with the field. For example, in the even-$n$ case, the term for $k=n$ gives the denominator $(a\chi_n\Omega_n + \omega)$, which is singular when $\Omega_n = -\omega/(a\chi_n)$. This is the familiar condition for the initial Unruh excitation, where the detector absorbs a quantum of energy from the vacuum fluctuations.

\subsubsection*{Field-Mediated Resonances}
For even-$n$ processes with $n>2$, a new class of resonances appears that is independent of $\omega$. For instance, in the four-photon case ($n=4$), the denominator for $k=3$ is simply $(a(\chi_3\Omega_3+\chi_4\Omega_4))$. This term leads to a resonance when $\Omega_4 = -\chi_3\chi_4\Omega_3$. This represents a purely field-mediated process where a pair of photons is created in a correlated state without direct energy exchange with the detector's internal levels. It suggests that the detector acts as a catalyst, enabling the Minkowski vacuum to radiate multi-particle states whose correlations are governed by the geometric properties of the accelerated frame itself.

\begin{figure}[ht]
    \centering
    \includegraphics[width=\textwidth]{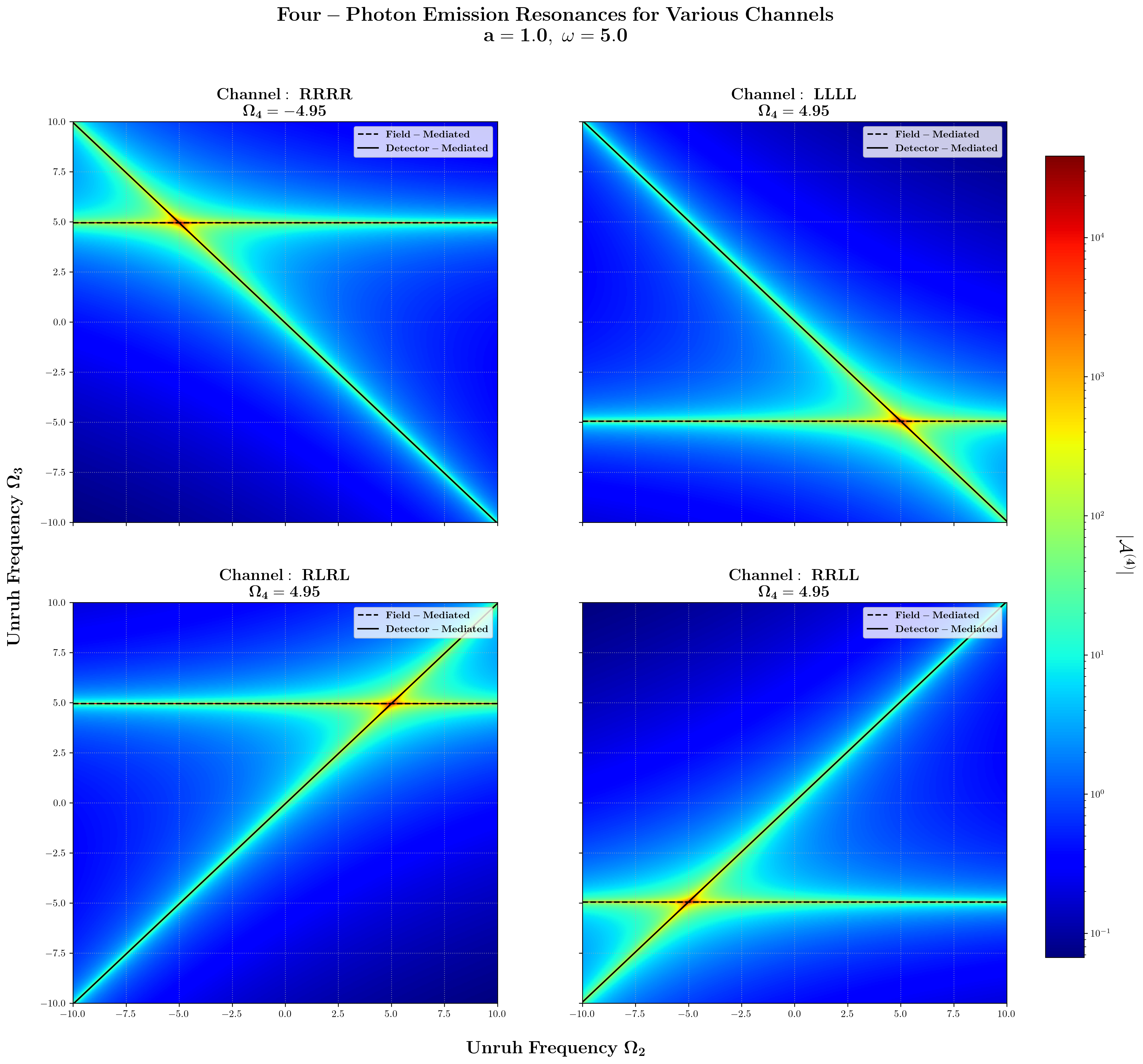}
    \caption{Density plots of the four-photon emission amplitude $|\mathcal{A}^{(4)}|$ for various directional channels. The axes represent the Unruh frequencies of two of the emitted photons. The bright lines indicate resonances. The solid lines are detector-mediated resonances, dependent on $\omega$, while the dashed lines are field-mediated resonances, independent of $\omega$. The different kinematic constraints for each channel are clearly visible.}
    \label{fig:multi_panel_resonances}
\end{figure}

These two resonance types are clearly visualized in Fig. \ref{fig:multi_panel_resonances}. The figure shows the four-photon emission amplitude for various directional channels. In each panel, the solid black line represents a detector-mediated resonance, whose position depends on $\omega$. The dashed black line represents a field-mediated resonance, whose position is independent of $\omega$. The intersection of these lines indicates a doubly resonant condition where the emission probability is maximally enhanced. The different orientations of these lines across the various channels (RRRR, LLLL, RLRL, RRLL) illustrate the rich kinematic structure of these higher-order processes.

\subsection{Thermal Nature and Detailed Balance}

The most striking feature of the odd-$n$ processes is the appearance of the explicit thermal factor $e^{-\pi\omega/2a}$ in the amplitude. This factor arises directly from the energy conservation condition, $\sum \chi_i\Omega_i = -\omega/a$, which is unique to processes that change the detector's state. The total probability for such a process, obtained by integrating the squared amplitude over all final state frequencies, is therefore proportional to the Boltzmann factor, $e^{-\pi\omega/a}$.

\begin{figure}[ht]
    \centering
    \includegraphics[width=0.7\textwidth]{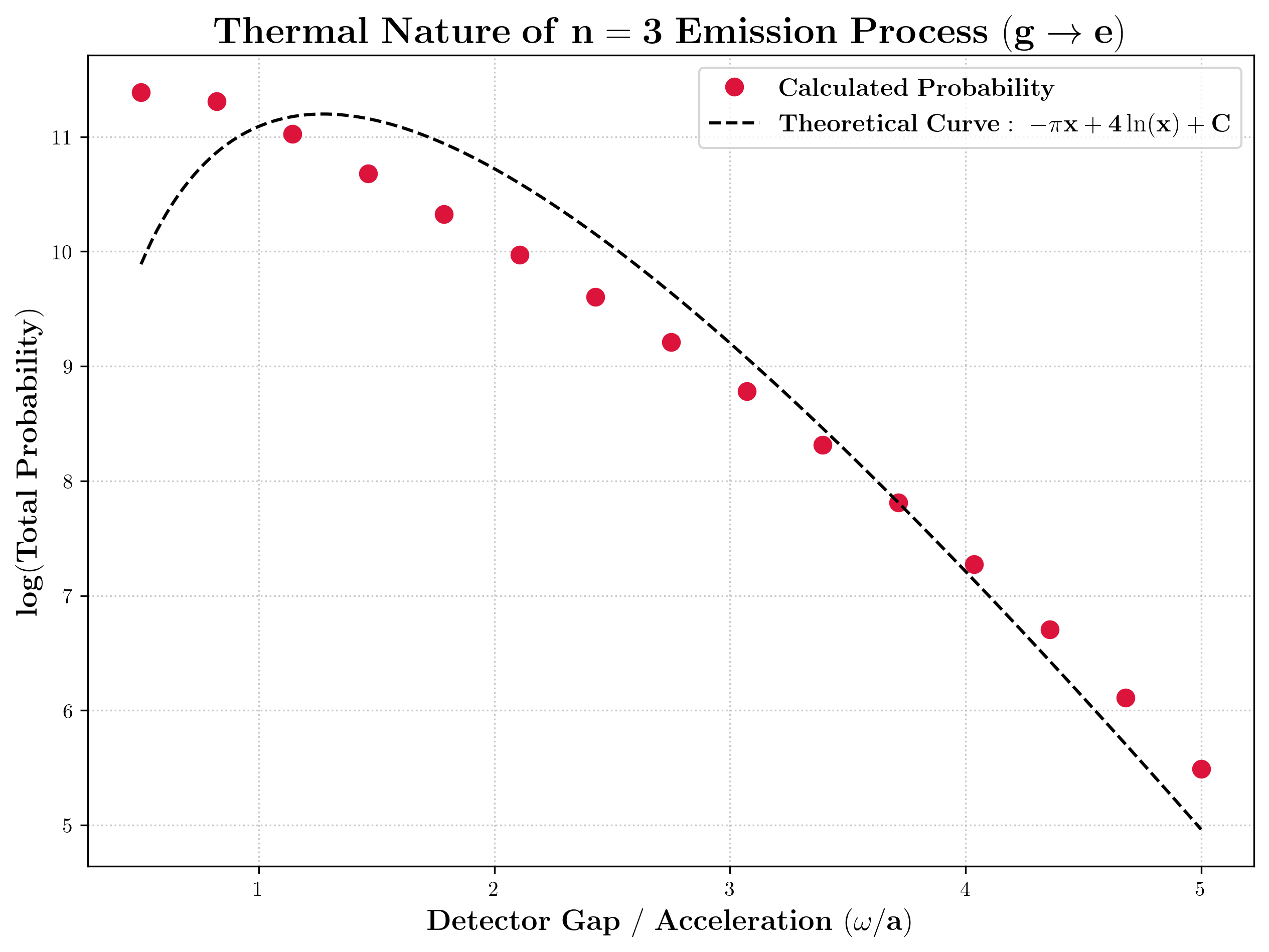}
    \caption{The logarithm of the total probability for the n=3 ($g \to e$) emission process versus the ratio of detector gap to acceleration, $\omega/a$. The red dots are the results of numerical integration of the squared amplitude. The dashed black line is the best-fit theoretical curve, which is dominated by the linear slope of $-\pi$ characteristic of the Unruh thermal factor. This provides strong visual confirmation of the Boltzmann-like suppression of excitation.}
    \label{fig:thermal_plot}
\end{figure}

This thermal behavior is confirmed numerically in Fig. \ref{fig:thermal_plot}. The figure plots the logarithm of the total probability for the n=3 ($g \to e$) emission process as a function of the ratio $\omega/a$. The calculated data points (red circles) are shown to fall along a theoretical curve (dashed black line) that is dominated by the exponential decay. The excellent agreement provides strong evidence that the n-th order calculation correctly captures the thermal nature of the Unruh effect. The slight deviation from a perfect straight line is due to additional power-law dependencies on $\omega/a$ from the pre-factors in the amplitude, which are captured by the theoretical curve of the form $-\pi x + C\ln(x) + \text{Constant}$.

\subsection{Multipartite Entanglement}
Finally, the delta function constraint ($\sum \chi_j \Omega_j = 0$ for even $n$, $\sum \chi_j \Omega_j = -\omega/a$ for odd $n$) imposes a strong correlation among all $n$ emitted photons. This generalizes the bipartite entanglement of the two-photon case to genuine multipartite entanglement for $n>2$. The structure of this entanglement is a direct imprint of the underlying correlations of the Minkowski vacuum, as probed by the higher-order interaction with the accelerated detector. For example, a four-photon RRLL channel ($A_{\Omega_1}^\dagger A_{\Omega_2}^\dagger B_{\Omega_3}^\dagger B_{\Omega_4}^\dagger$) would describe a state with four entangled particles distributed across both Rindler wedges and propagating in different directions, with a complex correlation structure dictated by the resonance conditions shown in Fig. \ref{fig:multi_panel_resonances}.

\section{Conclusion}

In this paper, we developed a comprehensive framework for $n^\text{th}$-order quantum processes involving a uniformly accelerated Unruh–DeWitt detector. By evaluating the time-ordered Dyson series exactly, we derived analytical formulas for the final $n$-photon quantum states for both even- and odd-order transitions. The Unruh-mode expansion is the key technical ingredient: when pulled back to the uniformly accelerated worldline, the field reduces to pure exponentials in proper time, allowing all $n$-vertex integrals to close into a delta function over chained denominators. \emph{This is precisely why we work in Unruh modes: their $w^{\pm i\Omega}$ structure is Rindler-like and maps along the trajectory to $e^{\mp i\chi a\Omega\,\tau}$, so each vertex contributes a pure exponential in $\tau$ and the full $n$-fold time integral becomes analytically tractable.}

Our results provide a multi-faceted confirmation and extension of the Unruh effect. For even-order processes, we identified new \emph{field-mediated} resonances (independent of $\omega$) in addition to the familiar \emph{detector-mediated} ones (dependent on $\omega$), showing how an accelerating detector can catalyze correlated multi-particle creation beyond simple excitation/de-excitation. For odd-order processes, the amplitudes acquire the explicit factor $e^{-\pi\omega/(2a)}$, so that probabilities carry the Boltzmann suppression $e^{-\pi\omega/a}$ characteristic of Unruh thermality. Most significantly, by comparing excitation ($g\!\to\!e$) and de-excitation ($e\!\to\!g$), we showed that detailed balance holds at arbitrary order, yielding the ratio $P(e\!\to\!g)/P(g\!\to\!e)=e^{2\pi\omega/a}$ at the Unruh temperature $T_U=a/(2\pi k_B)$.

The $n$-photon states exhibit genuine multipartite correlations: the delta-function constraints ($\sum \chi_j\Omega_j=0$ for even $n$, $\sum \chi_j\Omega_j=-\omega/a$ for odd $n$) enforce frequency-domain constraints that generalize the bipartite entanglement of the two-photon case to higher orders. The resonance geometry—intersections of detector- and field-mediated resonance lines—provides a clear, physically transparent organizing principle for these multipartite structures.

\paragraph*{Scope note: circular motion.}
Our analysis is intentionally restricted to \emph{uniform linear acceleration}. The Unruh-mode machinery hinges on the fact that, along a uniformly accelerated worldline, the pulled-back field becomes a sum of pure exponentials in proper time, which makes the $n$-fold time integrals analytic. In \emph{uniform circular motion} there is no canonical Unruh-mode basis aligned with proper time, and Minkowski plane waves on the circular worldline do not reduce to simple exponentials; a Jacobi–Anger decomposition leads instead to infinite harmonic sums that typically require numerical treatment. Moreover, the circular response is stationary but  non-Kubo–Martin–Schwinger (KMS) \cite{Kubo1957, Martin_Schwinger1959}  and generically non-Planckian, with at best regime-dependent, approximate thermal features (e.g., connections to Sokolov--Ternov in large-$\gamma$/small-gap regimes) \cite{Bell_Leinaas1987,Fewster2016Louko,Akhmedov_Singleton2007,Akhmedov_Singleton2008,Bunney2023Circularmotion}. We therefore do not extend our $n$-photon formulas to circular motion here, and we refrain from drawing quantitative conclusions for that case.

\paragraph*{Outlook.}
The formalism opens several directions: probing multipartite entanglement witnesses tailored to the resonance geometry; mapping how field-mediated ladders imprint on detector readouts; exploring non-thermal corrections at higher orders; and generalizing to more complex internal structures or curved backgrounds. A separate, numerically focused program could address uniform circular motion using harmonic expansions, with care to its non-KMS character and only approximate thermality. Ultimately, moving beyond leading order yields a sharper picture of how acceleration sculpts vacuum correlations and multi-quantum emission.

\section*{Acknowledgments}
I am grateful to Marlan Scully, Anatoly Svidzinsky, and Bill Unruh for illuminating discussions. This work was supported by the Robert A. Welch Foundation (Grant No. A-1261) and the National Science Foundation (Grant No. PHY-2013771).

\bibliographystyle{jhep}
\bibliography{UnruhRef} 
\end{document}